\documentstyle[psfig]{article}

\oddsidemargin=\evensidemargin
\newcommand{\sssection}[1]{%
     \subsubsection[#1]{\large{\it{#1}}}}

\newenvironment{myitemize}{\begin{itemize}
                \setlength{\itemsep}{-1 pt}}{\end{itemize}}

\newcommand{\dzero}     {D\O}
\newcommand{\rar}       {\rightarrow}
\newcommand{\rargap}    {\mbox{ $\rightarrow$ }}
\newcommand{\ttbar}     {\mbox{$t\bar{t}$}}
\newcommand{\bbbar}     {\mbox{$b\bar{b}$}}
\newcommand{\ppbar}     {\mbox{$p\bar{p}$}}
\newcommand{\qqbar}     {\mbox{$q\bar{q}$}}
\newcommand{\bbar}      {\mbox{$\bar{b}$}}
\newcommand{\Wbb}       {\mbox{$Wb\bar{b}$}}
\newcommand{\Wcc}       {\mbox{$Wc\bar{c}$}}

\newcommand{\deta}      {\mbox{$\eta^{\rm det}$}}
\newcommand{\mdeta}     {\mbox{$|\eta^{\rm det}|$}}
\newcommand{\met}       {\mbox{$\not\!\!E_T$}}
\newcommand{\metc}      {\mbox{$\not\!\!E_T^{\rm cal}$}}
\newcommand{\ejm}       {\mbox{$e$+jets/$\mu$}}
\newcommand{\mjm}       {\mbox{$\mu$+jets/$\mu$}}

\begin{document}

\begin{flushright}
{\dzero} Note 3817 ~~ UCR/{\dzero}/01-02

\vspace{-0.015in}
Fermilab-Conf-01/005-E

\vspace{-0.015in}
January 2001
\end{flushright}

\vspace{0.05 in}

\begin{center}

{\Large \bf A Search for Single Top Quark Production at {\dzero}} \\

\vspace{4mm}

A.P. HEINSON \\
Department of Physics, University of California \\
Riverside, CA 92521-0413, USA \\

\vspace{2mm}

FOR THE {\dzero} COLLABORATION \\

\end{center}

\begin{abstract}
We present details of a search for electroweak production of single
top quarks in the electron+jets and muon+jets decay channels. The
measurements use $\approx90$~pb$^{-1}$ of data from Run~1 of the
Fermilab Tevatron collider, collected at 1.8~TeV with the {\dzero}
detector. We use events that include a tagging muon, implying the
presence of a $b$ jet, to set an upper limit at the $95\%$ confidence
level on the cross section for the $s$-channel process
${\ppbar}{\rargap}tb+X$ of 39~pb. The upper limit for the $t$-channel
process ${\ppbar}{\rargap}tqb+X$ is 58~pb.
\end{abstract}

\section{Introduction}

The {\dzero} collaboration has recently published the results of a
search for single top quarks produced in association with a bottom
quark or a light quark and a low-$p_T$ $b$~quark~\cite{sintop-prd}.
The CDF collaboration has reported similar
measurements~\cite{cdfpaper}. These analyses search for two
independent modes that produce top quarks singly: the $s$-channel
process $q^{\prime}\bar{q}{\rar}tb$ with a predicted cross section of
{\mbox{$\sigma=0.75\pm0.12$~pb}~\cite{willenbrock-s}}; and the
$t$-channel process $q^{\prime}g{\rar}tqb$ with
{\mbox{$\sigma=1.47\pm0.22$~pb~\cite{willenbrock-t}}. These values
have been recently updated~\cite{willenbrock-s,willenbrock-t}. Events
are identified by the presence of one isolated electron or muon, and
missing transverse momentum assumed to be from the decay of a
$W$~boson to a lepton and neutrino. The events must also contain two
to four jets, with one or more having an associated muon to tag it as
a possible $b$~jet.

The published paper of the {\dzero} results contains a complete
summary of the analysis and details of the data and Monte Carlo event
samples and electron and muon identification criteria. Efficient
identification with a low fake rate is crucial to the success of the
search. At the start of the analysis, the dominant background in the
electron channel is from multijet events with a jet misidentified as
an electron, and in the muon channel the main background is from
events without a real muon from a $W$~boson decay.  An in-depth
discussion of the backgrounds and of the efficiencies for trigger
selection, particle identification, and cosmic ray rejection is
available in a conference paper~\cite{heinson-dpf}. Here, we focus on
the details of the event selections, and on the properties of the
final candidate events.

We use the notation ``$tb$'' to refer to both $t\bar{b}$ and the
charge-conjugate process $\bar{t}b$, and ``$tqb$'' for both
$tq\bar{b}$ and~$\bar{t}\bar{q}b$. The backgrounds referred to as
{\ttbar} and {\Wbb} are self-explanatory. The {\Wcc} background
includes all contributions from ${\ppbar}{\rar}W$+$c\bar{c},
W$+$c\bar{s}, W$+$\bar{c}s,$ and $W$+$s\bar{s}$. The $Wjj$ background
includes events with only $u$, $d$, or $g$ jets. The {\Wbb}, {\Wcc},
$Wjj$, $WW$, and $WZ$ sets are Monte Carlo samples used to cross check
the $W$+jets background, which is measured using data.

\section{Baseline Event Selections}

We apply a ``baseline'' set of event selections in order to choose all
possible candidates after triggering. The electron channel and muon
channel baseline samples are defined as those events which have at
least one isolated lepton of the type expected and two or more
jets. For data, the events must pass at least one of the Level 2
filters in the trigger and both of the Main Ring vetoes. Monte Carlo
(MC) events must pass at least one filter. The baseline selections are
given in Table~\ref{table1}. The effects of these extremely loose
selections on the data and MC events are shown in Table~\ref{table2}.

The baseline samples for QCD are defined slightly differently to the
others: the sample in the electron channel has at least three jets,
not two, and no isolated electron is required; the sample in the muon
channel has at least one nonisolated muon instead of the isolated one,
as well as the two jets.

\vspace{0.25in}
\footnoterule
\small
\noindent Invited plenary talk presented at the XVth International
Workshop on High Energy Physics and Quantum Field Theory, Tver,
Russia, 14th--20th September 2000.
\normalsize

\clearpage
\topmargin 0.5in
\footskip -1 in

Fewer muon channel events make it into the baseline samples than
electron channel ones because of the large difference in overall
lepton identification (ID) efficiencies. Of leptons that generate a
trigger, we reconstruct and identify only $\sim$0.6$\times$ as many
isolated muons as electrons.

\vspace{-0.1in}
\begin{table}[!h!tbp]
\begin{center}
\begin{tabular}{clccc}
\hline \hline
 & & & & \\ [-0.12 in]
\multicolumn{5}{c}{\large{Baseline Event Selections}}
\vspace{0.03 in}                                   \\
\hline
 Cut   &   &   &   &   Main Backgrounds            \\
  No.  & \multicolumn{1}{c}{Variable Definition} &  Variable Name
       &     Cutoff     &   Rejected               \\
\hline
 & & & & \\ [-0.12 in]
\multicolumn{5}{l}{\underline{Electron and Muon Channels}}
\vspace{0.04 in}                                   \\
     1     &  Pass the triggers and filters & &
           &  QCD                                  \\
           &  (and vetoes for data) & & &          \\
     2     &  Min. transverse energy of jet 1
           &  $E_T({\rm jet1})$
           &  $> 5$~GeV
           &  QCD, $W$+jets                        \\
     3     &  Min. transverse energy of jet 2
           &  $E_T({\rm jet2})$
           &  $> 5$~GeV
           &  QCD, $W$+jets                        \\
     4     &  Max. pseudorapidity of jet 1
           &  $|\eta^{\rm det}({\rm jet1})|$
           &  $< 4.0$
           &  QCD                                  \\
     5     &  Max. pseudorapidity of jet 2
           &  $|\eta^{\rm det}({\rm jet2})|$
           &  $< 4.0$
           &  QCD
\vspace{0.04 in}                                   \\
\multicolumn{5}{l}
      {\underline{Electron Channel Only}}
\vspace{0.04 in}                                   \\
     6     &  PELC that passes electron ID & &
           &  jets, photons                        \\
     7     &  Min. transverse energy of electron
           &  $E_T(e)$
           &  $> 20$~GeV
           &  QCD                                  \\
     8     &  Fiducial pseudorapidity of electron
           &  $|\eta^{\rm det}(e)|$
           &  $< 1.1, 1.5$--2.5
           &
\vspace{0.04 in}                                   \\
\multicolumn{5}{l}
      {\underline{Muon Channel Only}}
\vspace{0.04 in}                                   \\
     9     &  PMUO that passes isolated muon ID & &
           &  cosmics, patrec errors               \\
    10     &  Min. transverse momentum of muon
           &  $p_T(\mu)$
           &  $> 20$~GeV
           &  QCD                                  \\
    11     &  Fiducial pseudorapidity of muon
           &  $|\eta^{\rm det}(\mu)|$
           &  $< 1.7$  &
\vspace{0.04 in}                                   \\
\multicolumn{5}{l}
      {\underline{Tagging Muon}}
\vspace{0.04 in}                                   \\
    12     &  PMUO that passes tagging muon ID & &
           &  cosmics, patrec errors               \\
    13     &  Min. transverse momentum of muon
           &  $p_T(\mu)$
           &  $> 4$~GeV
           &  QCD                                  \\
    14     &  Fiducial pseudorapidity of muon
           &  $|\eta^{\rm det}(\mu)|$
           &  $< 1.7$  &
\vspace{0.04 in}                                   \\
\hline \hline
\end{tabular}
\caption[tab1]{The baseline event selection variables and cutoffs in
the electron and muon channels. A PELC is an energy cluster in the
calorimeter that has passed certain criteria in {\dzero}'s main event
reconstruction package RECO. A PMUO is a muon candidate from
RECO. ``patrec'' is short for ``pattern recognition''.}
\label{table1}
\end{center}
\end{table}

\vspace{-0.3in}
\begin{table}[!h!tbp]
\begin{center}
\begin{tabular}{lcccc}
\hline \hline
 & & & & \\ [-0.12 in]
\multicolumn{5}{c} {\large{Baseline Selection Efficiencies}}
\vspace{0.03 in}                                        \\
\hline
      & \multicolumn{2}{c} {\underline{Electron Channel}}
      & \multicolumn{2}{c} {\underline{Muon Channel}}   \\
Event & \multicolumn{2}{c} {$\%$ of Post-Trigger Sample}
      & \multicolumn{2}{c} {$\%$ of Post-Trigger Sample}\\
Type  &  Before $\mu$-Tag  &  After $\mu$-Tag
      &  Before $\mu$-Tag  &  After $\mu$-Tag           \\
\hline
 & & & & \\ [-0.12 in]
Signals         &        &         &        &           \\
~~MC $tb$       & $59\%$ & $8.2\%$ & $24\%$ & $3.5\%$   \\
~~MC $tqb$      & $62\%$ & $5.9\%$ & $27\%$ & $2.7\%$  
\vspace{0.02 in}                                        \\
Backgrounds     &        &         &        &           \\
~~MC {\ttbar}   & $71\%$ &$13.5\%$ & $28\%$ & $5.6\%$   \\
~~MC {\Wbb}     & $50\%$ & $4.7\%$ & $24\%$ & $2.2\%$   \\
~~MC {\Wcc}     & $55\%$ & $1.2\%$ & $29\%$ & $0.5\%$   \\
~~MC $Wjj$      & $51\%$ & $0.1\%$ & $26\%$ & $0.1\%$   \\
~~MC $WW$       & $62\%$ & $1.1\%$ & $20\%$ & $0.4\%$   \\
~~MC $WZ$       & $61\%$ & $2.2\%$ & $20\%$ & $0.7\%$   \\
~~QCD data      & $79\%$ & $0.4\%$ & $2.7\%$& $0.1\%$
\vspace{0.04 in}                                        \\
Signal data     &  $1\%$ &$0.01\%$ & $0.3\%$&$0.01\%$
\vspace{0.04 in}                                        \\
\hline
\hline
\end{tabular}
\caption[tab2]{For the electron and muon channels, the percentage of
post-trigger event samples which remain in the baseline samples
before and after applying the requirement for a tagging muon.}
\label{table2}
\end{center}
\end{table}

\clearpage
\topmargin 0.5in

\section{Loose Event Selections}

We apply a number of cleanup cuts to the baseline event samples in
order to remove misreconstructed events and those that have final
state objects in them which are not expected in the signals. These
cuts and their effects are described here.

\subsection{Loose Cuts for the Electron and Muon Channels}

\sssection{Extra Leptons and Photons}

We reject events from the baseline samples which have more than one
isolated lepton in them that passes the identification requirements. A
muon can have high $p_T$ ($> 20$~GeV) or low $p_T$
($4<p_T\leq20$~GeV). This cut is designed to remove $Z$, {\ttbar},
$WW$, and $WZ{\rargap}{\rm dileptons}$ backgrounds. It also rejects
some cosmic ray events in the muon channel. In addition, we remove
events containing one or more photons. This is intended to reject
${\ttbar}{\rar}ee$ and $Z{\rar} ee$ events where one of the electrons
has not had its track reconstructed (which happens for $\sim$10$\%$ of
fiducial electrons), $W\gamma$+jets events, events where there is a
bremsstrahlunged photon, and events where a jet fakes a photon.

Table~\ref{table3} shows the percentage of events in the tagged
baseline samples that fail each of these cuts either exclusively
(i.e., they fail exactly one of these cuts and pass all other loose
selections), or inclusively (i.e., they fail one of these cuts and any
other of the loose set of cuts). These four cuts reject $13.6\%$
($e$-channel) and $8.7\%$ ($\mu$-channel) of the {\ttbar} background
events that are not removed by any other cuts, while rejecting less
than $1\%$ of the $s$-channel signal events and only 1.9--2.7$\%$ of
the $t$-channel ones. Note that the $WW$ and $WZ$ MC samples do not
include dilepton decays, otherwise the rejection rates could be higher
than shown.

\begin{table}[!h!tbp]
\begin{tabular}{lcccccccc}
\hline \hline
 & & & & & & & & \\ [-0.12 in]
\multicolumn{9}{c} {\large{Effects of the Extra Object Rejection Cuts}}
\vspace{0.03 in}                                                               \\
\hline
  & \multicolumn{4}{c}{Electron Channel} & \multicolumn{4}{c}{Muon Channel}    \\
  &  Electron  &  High-$p_T \mu$  &  Low-$p_T \mu$  &  Photon
  &  Electron  &  High-$p_T \mu$  &  Low-$p_T \mu$  &  Photon                  \\
\hline
 & & & & & & & & \\ [-0.12 in]
               & \multicolumn{8}{c} {\underline{Fail Exclusively}}             \\
Signals        &       &       &       &       &       &       &       &       \\
~~MC $tb$      &  ---  &  ---  &  ---  &  0.4  &  ---  &  ---  &  0.1  &  0.1  \\
~~MC $tqb$     &  0.4  &  ---  &  0.3  &  2.0  &  0.4  &  ---  &  0.1  &  1.4  \\
Backgrounds    &       &       &       &       &       &       &       &       \\
~~MC {\ttbar}  &  3.4  &  3.3  &  1.6  &  5.3  &  3.9  &  1.0  &  0.7  &  2.9  \\
~~MC $WW$      &  ---  &  ---  &  ---  &  4.2  &  ---  &  ---  &  ---  &  ---  \\
~~MC $WZ$      &  ---  &  ---  &  ---  &  1.4  &  0.6  &  ---  &  ---  &  1.2  \\
~~QCD data     &  ---  &  ---  &  ---  &  ---  &  ---  &  ---  &  0.4  &  ---  \\
Signal data    &  ---  &  ---  &  ---  &  ---  &  ---  &  ---  &  ---  &  ---  \\
\vspace{0.04 in}
               & \multicolumn{8}{c} {\underline{Fail Inclusively}}             \\
Signals        &       &       &       &       &       &       &       &       \\
~~MC $tb$      &  ---  &  ---  &  ---  &  0.5  &  0.2  &  0.1  &  0.2  &  0.3  \\
~~MC $tqb$     &  0.5  &  ---  &  0.4  &  2.7  &  0.6  &  ---  &  0.2  &  3.2  \\
Backgrounds    &       &       &       &       &       &       &       &       \\
~~MC {\ttbar}  &  4.3  &  5.1  &  2.2  &  8.0  &  8.5  &  2.2  &  1.4  &  7.7  \\
~~MC $WW$      &  ---  &  ---  &  ---  &  5.6  &  ---  &  ---  &  ---  &  ---  \\
~~MC $WZ$      &  0.7  &  0.3  &  ---  &  1.4  &  2.4  &  ---  &  0.6  &  4.2  \\
~~QCD data     &  ---  &  ---  &  ---  &  ---  &  ---  &  ---  &  6.9  &  0.1  \\
Signal data    &  ---  &  ---  &  ---  &  ---  &  ---  &  ---  &  1.8  &  ---
\vspace{0.04 in}                                                               \\
\hline
\hline
\end{tabular}
\caption[tab3]{Percentages of the tagged baseline event sets which
fail each of the extra lepton or photon vetoes exclusively or
inclusively. The $W$+jets MC samples (not shown) are negligibly
affected.}
\label{table3}
\end{table}

\clearpage

\sssection{Mismeasured Jets}

We check the quality of every jet in the data with $E_T > 5$~GeV and
${\mdeta} < 4.0$. We do not apply these cuts to MC since the details
of the jets are not modeled well enough and the sources of noise are
not present. Instead, we correct for the small loss in efficiency. The
quality checks are:

\begin{myitemize}
\item Fraction of $E_T$ in the electromagnetic calorimeter layers
$(F(E_T^{\rm EM})) < 0.9$
\item $-0.05 <$ Fraction of $E_T$ in the coarse hadronic calorimeter
layers $(F(E_T^{\rm CH})) < 0.5$
\item Ratio of $E_T$'s of hottest cell in jet to next-hottest cell
$(R_{\rm Hotcell}) < 10$
\end{myitemize}

If any jet in an event fails any of these requirements, we call it a
``bad jet'' and discard the event, since the $E_T$ of the jet cannot
be relied upon. The fraction of $E_T$ in the coarse hadronic
calorimeter layers can go slightly negative because corrections for
hot cells have been made in the reconstruction package before the
cells are clustered into jets, and occasionally more energy has been
subtracted from an individul cell's energy than was necessary.
Table~\ref{table4} shows the exclusive and inclusive percentages of
tagged baseline events which fail these cuts.

\begin{table}[!h!tbp]
\begin{center}
\begin{tabular}{lcccccc}
\hline \hline
 & & & & & & \\ [-0.12 in]
\multicolumn{7}{c} {\large{Effects of the Jet Quality Cuts}}
\vspace{0.03 in}                                             \\
\hline
Event  & \multicolumn{3}{c} {Electron Channel}
       & \multicolumn{3}{c} {Muon Channel}                   \\
Type   & $F(E_T^{\rm EM})$ & $F(E_T^{\rm CH})$ & $R_{\rm Hotcell}$
       & $F(E_T^{\rm EM})$ & $F(E_T^{\rm CH})$ & $R_{\rm Hotcell}$ \\
\hline
 & & & & & & \\ [-0.12 in]
 & \multicolumn{6}{c} {\underline{Fail Exclusively}}         \\
 & & & & & & \\ [-0.12 in]
QCD data     &  0.7  &  0.4  &  0.1  &  0.2  &  0.5  &  ---  \\
Signal data  &  0.9  &  1.7  &  ---  &  0.9  &  ---  &  ---  \\
 & \multicolumn{6}{c} {\underline{Fail Inclusively}}         \\
 & & & & & & \\ [-0.12 in]
QCD data     &  5.1  &  2.3  &  0.6  &  4.3  &  3.0  &  0.4  \\
Signal data  &  2.6  &  2.6  &  ---  &  2.7  &  2.7  &  0.9
\vspace{0.04 in}                                             \\
\hline
\hline
\end{tabular}
\vspace{0.2 in}
\caption[tab4]{Percentage of the tagged baseline event samples
which fail each jet quality requirement exclusively or inclusively.}
\label{table4}
\end{center}
\end{table}

\clearpage

\sssection{Jet $E_T$ and {\mdeta} Cuts}

We now apply some cuts on the jets designed to keep as much signal
acceptance as possible while rejecting some obvious backgrounds in the
baseline samples. The requirements on the jets are:

\begin{myitemize}
\item $E_T({\rm jet1}) > 15$~GeV
\item $E_T({\rm jet2}) > 10$~GeV
\item $|\eta^{\rm det}({\rm jet1})| < 3.0$
\end{myitemize}

\noindent After we have made these demands, we require:

\begin{myitemize}
\item $2 \leq n_{\rm jets} \leq 4$
\end{myitemize}

Table~\ref{table5} shows the percentages of tagged baseline events
which fail the minimum $E_T$, maximum {\mdeta}, and multiplicity
requirement, either exclusively (fail exactly one of these cuts and
pass all other loose selections) or inclusively (fail one of these
cuts plus any other of the loose cuts).

\begin{table}[!h!tbp]
\begin{center}
\begin{tabular}{lcccccccc}
\hline \hline
 & & & & & & & & \\ [-0.12 in]
\multicolumn{9}{c} {\large{Effects of the Jet Kinematics Cuts}}
\vspace{0.03 in}                                                 \\
\hline
  & \multicolumn{4}{c} {Electron Channel}
  & \multicolumn{4}{c} {Muon Channel}                            \\
  &  $E_T^{\rm min}$             &  $|\eta^{\rm det}_{\rm max}|$
  &  $n_{\rm jets}^{\rm \:min}$  &  $n_{\rm jets}^{\rm \:max}$
  &  $E_T^{\rm min}$             &  $|\eta^{\rm det}_{\rm max}|$
  &  $n_{\rm jets}^{\rm \:min}$  &  $n_{\rm jets}^{\rm \:max}$   \\
\hline
 & & & & & & & & \\ [-0.12 in]
  & \multicolumn{8}{c} {\underline{Fail Exclusively}}                            \\
Signals          &       &       &       &       &       &       &       &       \\
~~MC $tb$        &  0.6  &  ---  &       &  ---  &  0.3  &  ---  &       &  ---  \\
~~MC $tqb$       &  2.3  &  0.2  &       &  0.3  &  1.2  &  0.2  &       &  0.1  \\
Backgrounds      &       &       &       &       &       &       &       &       \\
~~MC {\ttbar}    &  4.4  &  0.1  &       & 21.8  &  3.7  &  0.1  &       & 14.5  \\
~~MC {\Wbb}      &  0.4  &  ---  &       &  ---  &  ---  &  ---  &       &  ---  \\
~~MC {\Wcc}      &  ---  &  ---  &       &  ---  &  ---  &  ---  &       &  ---  \\
~~MC $Wjj$       &  ---  &  ---  &       &  ---  &  ---  &  ---  &       &  ---  \\
~~MC $WW$        &  4.9  &  ---  &       &  7.6  &  5.6  &  ---  &       &  1.1  \\
~~MC $WZ$        &  4.8  &  0.3  &       &  3.4  &  3.6  &  ---  &       &  0.6  \\
~~QCD data       &  0.4  &  ---  &       &  1.3	 &  2.4  &  ---  &       &  0.7  
\vspace{0.04 in}                                                                 \\
Signal data      &  1.7  &  ---  &       &  0.9	 &  6.4  &  ---  &       &  0.9  
\vspace{0.04 in}                                                                 \\
  & \multicolumn{8}{c} {\underline{Fail Inclusively}}                            \\
Signals          &       &       &       &       &       &       &       &       \\
~~MC $tb$        &  4.9  &  0.1  &  4.3  &  ---  &  4.4  &  0.1  &  3.9  &  ---  \\
~~MC $tqb$       &  6.5  &  1.1  &  4.0  &  0.3  &  6.5  &  2.0  &  4.4  &  0.1  \\
Backgrounds      &       &       &       &       &       &       &       &       \\
~~MC {\ttbar}    &  9.2  &  1.2  &  1.1  & 27.5  & 10.9  &  1.4  &  0.9  & 27.1  \\
~~MC {\Wbb}      & 10.6  &  0.1  & 10.0  &  ---  & 11.5  &  0.1  & 11.4  &  ---  \\
~~MC {\Wcc}      & 13.3  &  0.6  & 13.3  &  ---  & 10.6  &  1.2  & 10.6  &  ---  \\
~~MC $Wjj$       & 23.3  &  ---  & 23.3  &  ---  & 31.6  &  3.6  & 31.6  &  ---  \\
~~MC $WW$        & 10.4  &  ---  &  5.6  &  8.6  & 13.5  &  1.1  &  3.4  &  4.5  \\
~~MC $WZ$        &  9.9  &  0.7  &  4.1  &  3.4  & 11.5  &  1.2  &  6.1  &  2.4  \\
~~QCD data       & 15.4  &  2.2  &  2.1  &  6.6	 & 20.7  &  2.3  &  5.0  &  4.0  
\vspace{0.04 in}                                                                 \\
Signal data      & 29.3  &  5.2  & 16.4  &  2.6  & 44.5  &  0.9  & 20.9  &  5.5  
\vspace{0.04 in}                                                                 \\
\hline
\hline
\end{tabular}
\caption[tab5]{Percentages of tagged baseline events which fail
each of the jet requirements exclusively or inclusively.  (Note that
events cannot fail the $n_{\rm jets}^{\rm min}$ cut exclusively, since
it forms part of the baseline requirements.)}
\label{table5}
\end{center}
\end{table}

\clearpage

\sssection{Missing Transverse Energy}

There is a neutrino in each of our signal events from the decay of the
$W$~boson from the top quark decay, with an average $E_T$ at the
parton level of $\sim$48~GeV. Therefore, we make the following
requirements of our events:

\begin{myitemize}
\item ${\metc} > 15$~GeV
\item ${\met} > 15$~GeV
\end{myitemize}

{\metc} is the vector transverse energy imbalance in the calorimeters
before correcting for energy carried away by any muons in the
event. {\met} is the missing transverse energy after making such
corrections.

Events with less than 15~GeV of {\met} (or {\metc}) are usually QCD
multijet events where there is a fake electron or fake isolated muon,
and where the {\met} is a fluctuation from the decay of a $b$~hadron
into a muon and its associated neutrino, or where one or more objects
in the event has been mismeasured, thus generating fake {\met}.

The choice of {\met} threshold of 15~GeV is a balance between two
competing issues. We could increase the signal acceptance without
much increase in the background by lowering the cut to
10~GeV. However, this then leaves less data below the {\met} threshold
with which to measure the probabilities for a jet to fake an electron
and for a nonisolated muon to fake an isolated one, which leads to
larger errors in the final result. In Run~2, a shortage of data will
no longer be a problem (and the fake electron probability should be
much smaller with the addition of a 2~T central solenoid magnet), and
therefore a lower {\met} threshold may be advantageous.

The percentages of tagged baseline events which fail the requirements
on {\metc} and {\met} are shown in Table~\ref{table6}.  It should be
noted that not many electron channel events fail this cut exclusively
because there is a special set of cuts (``mismeasured {\met}'',
described next) that is highly correlated with the {\metc} and {\met}
cuts.

\begin{table}[!h!tbp]
\begin{center}
\begin{tabular}{lcccc}
\hline \hline
 & & & & \\ [-0.12 in]
\multicolumn{5}{c}
{\large{Effects of the Missing Transverse Energy Cuts}}
\vspace{0.03 in}                                 \\
\hline
Event & \multicolumn{2}{c} {Electron Channel}
      & \multicolumn{2}{c} {Muon Channel}        \\
Type  &  Fail Exclusively  &  Fail Inclusively
      &  Fail Exclusively  &  Fail Inclusively   \\
\hline
 & & & & \\ [-0.12 in]
Signals        &       &       &        &        \\
~~MC $tb$      &  0.2  &  5.1  &   3.8  &   8.8  \\
~~MC $tqb$     &  0.2  &  5.6  &   2.7  &   7.2
\vspace{0.02 in}                                 \\
Backgrounds    &       &       &        &        \\
~~MC {\ttbar}  &  ---  &  5.9  &   1.5  &   7.6  \\
~~MC {\Wbb}    &  0.1  &  8.0  &   5.3  &  14.0  \\
~~MC {\Wcc}    &  ---  &  5.2  &   5.3  &  13.5  \\
~~MC $Wjj$     &  ---  &  2.3  &   5.3  &  10.5  \\
~~MC $WW$      &  ---  &  4.9  &   4.5  &   9.0  \\
~~MC $WZ$      &  ---  &  6.5  &   1.8  &   9.7  \\
~~QCD data     &  0.1  & 63.7  &  15.6  &  58.1
\vspace{0.04 in}                                 \\
Signal data    &  ---  & 63.8  &   2.7  &  25.5
\vspace{0.04 in}                                 \\
\hline
\hline
\end{tabular}
\caption[tab6]{Percentages of tagged baseline events which fail the
minimum {\metc} and {\met} requirements exclusively or inclusively.}
\label{table6}
\end{center}
\end{table}

\clearpage

\sssection{Mismeasured Missing Transverse Energy}

Some of the objects in the events are somewhat mismeasured, which
leads to fake missing transverse energy aligned with or back-to-back
with the object. There are many events near the ${\met} > 15$~GeV
threshold where the {\met} is back-to-back with the electron or
isolated muon. Events with mismeasured {\met} can also have it aligned
or anti-aligned with a jet. We therefore implement triangular-shaped
cuts in the ($\Delta\phi,{\met}$) plane to remove these events.

The cuts are defined as follows. We keep events if:

\begin{myitemize}
\item $(20/\pi)\times \Delta\phi(e,{\met}) - {\met} < 0.0$
\item $(20/\pi)\times \Delta\phi(\rm{jet}{\it i},{\met}) - {\met}
< 0$, for $i$ = 1,2,3,4 in the electron channel
\item $(20/\pi)\times \Delta\phi(\rm{jet}{\it i},{\met}) + {\met}
> 20$, for $i$ = 1,2,3,4 in the electron channel
\item $(240/\pi)\times \Delta\phi({\rm isol~}\mu,{\met}) - {\met}
< 190$
\end{myitemize}

The effects of these cuts on data and MC signals and backgrounds are
shown in Table~\ref{table7}. Many of the problems observed with the
jets are for ones in the intercryostat regions
($-1.4\leq{\deta}\leq-0.8$ and $0.8\leq{\deta}\leq1.4$). For
electrons, most of them are probably jets misidentified as electrons,
and as such, a different type of algorithm has been used to
reconstruct the energy than is appropriate. The $E_T$ found with the
cell-clustering algorithm and with a jet-cone algorithm is not the
same, and this can create a small amount of false missing transverse
energy. It can be seen from the table that the cuts are quite powerful
at rejecting QCD multijet background in the electron channel. In the
muon channel, the cuts do not need to be very tight because the QCD
{\bbbar} background there is very small, and it is better to keep
signal acceptance than reject more background in this case.

\begin{table}[!h!tbp]
\begin{center}
\begin{tabular}{lcccc}
\hline \hline
 & & & & \\ [-0.12 in]
\multicolumn{5}{c}
{\large{Effects of the Missing Transverse Energy ``Triangle'' Cuts}}
\vspace{0.03 in}                                 \\
\hline
Event & \multicolumn{2}{c} {Electron Channel}
      & \multicolumn{2}{c} {Muon Channel}        \\
Type  &  Fail Exclusively  &  Fail Inclusively
      &  Fail Exclusively  &  Fail Inclusively   \\
\hline
 & & & & \\ [-0.12 in]
Signals        &       &        &       &        \\
~~MC $tb$      &  2.0  &   6.9  &  1.5  &   4.3  \\
~~MC $tqb$     &  2.1  &   7.6  &  1.1  &   4.7
\vspace{0.02 in}                                 \\
Backgrounds    &       &        &       &        \\
~~MC {\ttbar}  &  1.3  &   8.8  &  0.4  &   3.4  \\
~~MC {\Wbb}    &  3.4  &  11.5  &  1.4  &   6.3  \\
~~MC {\Wcc}    &  2.8  &   8.0  &  0.0  &   5.3  \\
~~MC $Wjj$     &  2.3  &   4.7  &  5.3  &   5.3  \\
~~MC $WW$      &  3.5  &   9.7  &  5.6  &  11.2  \\
~~MC $WZ$      &  2.7  &   9.2  &  0.6  &   3.6  \\
~~QCD data     &  9.7  &  80.3  &  1.5  &  24.1
\vspace{0.04 in}                                 \\
Signal data    &  6.0  &  74.1  &  0.9  &  14.5
\vspace{0.04 in}                                 \\
\hline
\hline
\end{tabular}
\caption[tab7]{Percentages of tagged baseline events which have
an object aligned with or back-to-back to the {\met}, and the {\met}
is low, caused by a mismeasurement of the object's $E_T$.}
\label{table7}
\end{center}
\end{table}

\clearpage

\sssection{Mismeasured Tagging Muon Transverse Momentum}

To avoid problems with a neural network search (in progress), we
remove events if the tagging muon $p_T$ has been severely
mismeasured. We keep tagged events if:

\begin{myitemize}
\item $p_T({\rm tag~} \mu) < 500$~GeV
\end{myitemize}

\noindent Table~\ref{table8} shows the percentage of tagged baseline
events removed by this cut.

\begin{table}[!h!tbp]
\begin{center}
\begin{tabular}{lcccc}
\hline \hline
 & & & & \\ [-0.12 in]
\multicolumn{5}{c}
{\large{Effects of the Tagging Muon Maximum $p_T$ Cut}}
\vspace{0.03 in}                                 \\
\hline
Event & \multicolumn{2}{c} {Electron Channel}
      & \multicolumn{2}{c} {Muon Channel}        \\
Type  &  Fail Exclusively  &  Fail Inclusively
      &  Fail Exclusively  &  Fail Inclusively   \\
\hline
 & & & & \\ [-0.12 in]
Signals          &       &       &       &       \\
~~MC $tb$        &  0.2  &  0.2  &  0.2  &  0.3  \\
~~MC $tqb$       &  0.1  &  0.1  &  0.2  &  0.2  \\
Backgrounds      &       &       &       &       \\
~~MC {\ttbar}    &  0.1  &  0.3  &  ---  &  0.2  \\
~~MC {\Wbb}      &  ---  &  ---  &  ---  &  ---  \\
~~MC {\Wcc}      &  0.3  &  0.3  &  ---  &  ---  \\
~~MC $Wjj$       &  ---  &  ---  &  ---  &  ---  \\
~~MC $WW$        &  ---  &  ---  &  ---  &  ---  \\
~~MC $WZ$        &  ---  &  ---  &  ---  &  0.6  \\
~~QCD data       &  0.1  &  0.1  &  ---  &  0.1
\vspace{0.04 in}                                 \\
Signal data      &  0.9  &  0.9  &  ---  &  ---
\vspace{0.04 in}                                 \\
\hline
\hline
\end{tabular}
\caption[tab8]{Percentages of tagged baseline events which have
a badly mismeasured tagging muon.}
\label{table8}
\end{center}
\end{table}

\clearpage

\subsection{Additional Loose Cuts for the Muon Channel}

The muon channel needs several cleanup cuts which are not applicable
to the electron channel because there are more problems with isolated
muon reconstruction than with electrons, and because there are two
muons in the final state for tagged events, which give rise to
additional sources of background.

\sssection{Misreconstructed Isolated Muons}

When a muon has very high momentum, its track is not bent much in the
toroid, and its $p_T$ can be reconstructed to have an arbitrarily high
value. This in itself is not a problem, or an indication that the
reconstructed muon does not refer directly to a real high-$p_T$
muon. However, when it occurs, fake missing transverse energy is
generated in the event back-to-back with the muon. Since we use {\met}
to characterize events and to separate signal from background, if it
is corrupted, we can no longer be sure about the kinematics of the
event and it is best to reject such events.  Therefore, we keep events
only if:

\begin{myitemize}
\item $p_T({\rm isol} \mu) < 250$~GeV or ${\met} < 250$~GeV
\end{myitemize}

That is, we reject events where both $p_T^{\mu}$ and {\met} are high.
There is no explicit demand that the muon and {\met} be back-to-back
since almost every event which has high muon $p_T$ and high {\met} has
them back-to-back. They are also extremely correlated in magnitude,
because almost all of the {\met} is being generated by the
mismeasurement of the muon $p_T$. (There are a handful of events where
the {\met} is extremely high and the muon is properly measured; they
are also rejected by this cut.) The results are shown in
Table~\ref{table9}. For tagging muons, we do not see this problem and
therefore we do not need to apply a similar cut.

\begin{table}[!h!btp]
\begin{center}
\begin{tabular}{lcc}
\hline \hline
 & & \\ [-0.12 in]
\multicolumn{3}{c}
{\large{Effects of the Mismeasured Isolated Muon Cut}}
\vspace{0.03 in}                                      \\
\hline
Event Type  &  Fail Exclusively  &  Fail Inclusively  \\
\hline
 & & \\ [-0.12 in]
Signals          &        &        \\
~~MC $tb$        &   0.1  &   0.6  \\
~~MC $tqb$       &   0.2  &   0.8
\vspace{0.02 in}                   \\
Backgrounds      &        &        \\
~~MC {\ttbar}    &   0.1  &   1.0  \\
~~MC {\Wbb}      &   0.3  &   0.7  \\
~~MC {\Wcc}      &   ---  &   1.2  \\
~~MC $Wjj$       &   ---  &   ---  \\
~~MC $WW$        &   ---  &   ---  \\
~~MC $WZ$        &   ---  &   0.6  \\
~~QCD data       &   0.1  &   2.5
\vspace{0.04 in}                   \\
Signal data      &   ---  &   2.7
\vspace{0.04 in}                   \\
\hline
\hline
\end{tabular}
\caption[tab9]{Percentages of tagged baseline muon events which have
a very high-$p_T$ isolated muon and very high {\met}. When this
occurs, they are back-to-back and highly correlated in magnitude.}
\label{table9}
\end{center}
\end{table}

\clearpage

\sssection{Mismeasured Isolated Muon Transverse Momentum}

To avoid problems with a neural network search (in progress), we
remove events if the isolated muon $p_T$ has been severely
mismeasured. We keep muon channel events if:

\begin{myitemize}
\item $p_T({\rm isol} \mu) < 500$~GeV
\end{myitemize}

Table~\ref{table10} shows the percentage of tagged baseline events
removed by this cut. No events are removed solely because of it; it
was applied as a precautionary measure to protect the neural network
analysis.

\begin{table}[!h!tbp]
\begin{center}
\begin{tabular}{lc}
\hline \hline
 & \\ [-0.12 in]
\multicolumn{2}{c}{\large{Effects of the Isolated Muon Max $p_T$ Cut}}
\vspace{0.03 in}                 \\
\hline
Event Type  &  Fail Inclusively  \\
\hline
 & \\ [-0.12 in]
Signals          &       \\
~~MC $tb$        &  0.4  \\
~~MC $tqb$       &  0.5
\vspace{0.02 in}         \\
Backgrounds      &       \\
~~MC {\ttbar}    &  0.5  \\
~~MC {\Wbb}      &  0.1  \\
~~MC {\Wcc}      &  0.6  \\
~~MC $Wjj$       &  ---  \\
~~MC $WW$        &  ---  \\
~~MC $WZ$        &  0.6  \\
~~QCD data       &  1.4
\vspace{0.04 in}         \\
Signal data      &  2.7
\vspace{0.04 in}         \\
\hline
\hline
\end{tabular}
\caption[tab10]{Percentages of tagged baseline events which have a
badly mismeasured isolated muon. No events fail this cut and no other
cuts.}
\label{table10}
\end{center}
\end{table}

\clearpage

\sssection{Cosmic Rays}

In the muon channel, there is a significant residual contamination
from cosmic rays, even after all other loose selections have been
applied. Figure~\ref{fig1} shows $\Delta\phi(\rm{isol~} \mu,\rm{tag~}
\mu)$, the opening angle in the transverse plane between the isolated
and tagging muons in the event. In the data, there is a large peak
when the muons are back-to-back, indicative of cosmic ray
contamination. There is no matching peak in the opening angle in
{\deta} between the muons because the cosmic rays do not necessarily
pass through the {\dzero} detector at the same $z$ position as the
primary vertex. The pattern recognition algorithm tends to drop some
of the muon chamber hits from the tracks and pull the tracks in the
$r$-$z$ plane so that both halves of the track appear to originate
from the primary vertex. After track fitting, the tracks are no longer
sufficiently back-to-back in {\deta} to cause a peak in the
distribution. Since the peak in the $\Delta\phi$ distribution is sharp
and cannot be gotten rid of with any sophisticated examination of the
hits used on tracks, or muon $p_T$ difference, for example (and we
tried many things, since this cut has such a detrimental effect on
signal acceptance), we simply apply a cut on the $\Delta\phi$
distribution to remove this background. It is important to reject
these events from the analysis since they are not included in our
background model, and consequently they significantly degrade the
search sensitivity in the muon channel.

We keep events only if they have:

\begin{myitemize}
\item $\Delta\phi(\rm{isol} \mu,\rm{tag} \mu) < 2.4$~radians
\end{myitemize}

For the QCD background events, what is shown in Fig.~\ref{fig1} as
isolated muons is actually nonisolated ones. The peak near zero for
this data is therefore from double-tagged jets.

\vspace{0.2 in}
\begin{figure}[!h!tbp]
\centerline{\protect\psfig{figure=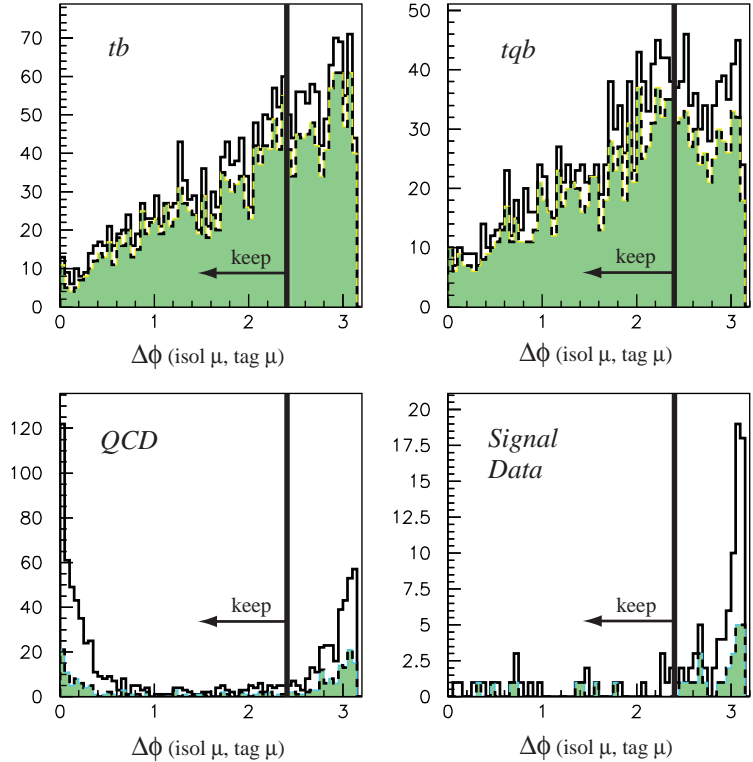,height=4in}}
\caption[fig1]{Distributions of the opening angle between the
isolated and tagging muon. The open histograms show the tagged
baseline samples,and the shaded histograms show the tagged loose set
of events. Cosmic ray contamination is seen in the data, but not in
Monte Carlo signal events.}
\label{fig1}
\end{figure}

We show the effects of the $\Delta\phi$ cut in Table~\ref{table11}.

\clearpage

\begin{table}[!h!tbp]
\begin{center}
\begin{tabular}{lcc}
\hline \hline
 & & \\ [-0.12 in]
\multicolumn{3}{c}{\large{Effects of the Cosmic Ray Cut}}
\vspace{0.03 in}                                      \\
\hline
Event Type  &  Fail Exclusively  &  Fail Inclusively  \\
\hline
 & & \\ [-0.12 in]
Signals          &                 \\
~~MC $tb$        &  33.5  &  39.9  \\
~~MC $tqb$       &  26.3  &  34.2
\vspace{0.02 in}                   \\
Backgrounds      &        &        \\
~~MC {\ttbar}    &  16.1  &  34.8  \\
~~MC {\Wbb}      &  31.3  &  41.8  \\
~~MC {\Wcc}      &  31.2  &  42.9  \\
~~MC $Wjj$       &  26.3  &  42.1  \\
~~MC $WW$        &  23.6  &  33.7  \\
~~MC $WZ$        &  31.5  &  44.8  \\
QCD data         &  13.2  &  39.5
\vspace{0.04 in}                   \\
Signal data      &  23.6  &  73.6
\vspace{0.04 in}                   \\
\hline
\hline
\end{tabular}
\caption[tab11]{Percentages of baseline tagged muon channel events
which have the isolated muon and a tagging muon back-to-back in
$\Delta\phi$.}
\label{table11}
\end{center}
\end{table}


\subsection{Summary of the Loose Event Selections}

Table~\ref{table12} summarizes the loose cuts used in this
analysis. These event selection requirements are applied to the
baseline samples of events, and are designed to keep as much of the
single top quark signals as possible while rejecting obvious nonsignal
events, in preparation for the final tight set of cuts. The
efficiencies of the loose selections are shown in Table~\ref{table13}.

After the baseline selections, there are 116 candidates in the
electron channel and 110 in the muon channel. These are almost all QCD
multijet events with a fake electron, or cosmic ray and fake
isolated-muon events. After the loose event selections, there remain
21 candidates in the electron channel and 8 in the muon channel; still
just over half are fake-lepton events. The percentage of these events
which are expected to be from single top quark production ($s$-channel
and $t$-channel combined) has improved from $0.53\%$ in the electron
channel after the baseline selections to $2.5\%$ after the loose
selections, and in the muon channel the percentage of signal has
increased from $0.43\%$ to $3.0\%$.

\begin{table}[!h!tbp]
\begin{center}
\begin{tabular}{clccc}
\hline \hline
 & & & & \\ [-0.12 in]
\multicolumn{5}{c} {\large{Loose Event Selections}}
\vspace{0.03 in}                                      \\
\hline
 Cut   &   &   &   &   Main Backgrounds            \\
  No.  & \multicolumn{1}{c}{Variable Definition} &  Variable Name
       &     Cutoff     &   Rejected               \\
\hline
 & & & & \\ [-0.12 in]
\multicolumn{5}{l}{\underline{Electron and Muon Channels}}
\vspace{0.04 in}                                   \\
     1     &  No extra electrons
           &  $n_e$
           &  $= 1$ or $= 0$
           &  {\ttbar}, $WZ$, $WW$                 \\
     2     &  No extra muons (low or high $p_T$)
           &  $n_{{\rm isol}\mu}$
           &  $= 0$ or $= 1$
           &  {\ttbar}, cosmics, $WZ$, $WW$        \\
     3     &  No photons
           &  $n_{\gamma}$
           &  $= 0$
           &  {\ttbar}, $WZ$, $WW$                 \\
     4     &  No bad jets
           &  $F(E_T^{\rm EM})$
           &  $< 0.9$
           &  Mismeasured events                   \\
         & &  $F(E_T^{\rm CH})$
           &  $> -0.05$, $<0.5$  &                 \\
         & &  $R_{\rm Hotcell}$
           &  $< 10$             &                 \\
     5     &  Min. transverse energy of jet1
           &  $E_T({\rm jet1})$
           &  $> 15$~GeV
           &  $W$+jets, QCD                        \\
     6     &  Min. transverse energy of jet2
           &  $E_T({\rm jet2})$
           &  $> 10$~GeV
           &  $W$+jets, QCD                        \\
     7     &  Max. pseudorapidity of jet1
           &  $|\eta^{\rm det}({\rm jet1})|$
           &  $< 3.0$
           &  $Wjj$, QCD                           \\
     8     &  Max. pseudorapidity of jet2
           &  $|\eta^{\rm det}({\rm jet2})|$
           &  $< 4.0$
           &  $Wjj$, QCD                           \\
     9     &  Minimum number of jets
           &  $n_{\rm jets}$
           &  $\ge 2$
           &  $W$+jets, $WW$, $WZ$                 \\
    10     &  Maximum number of jets
           &  $n_{\rm jets}$
           &  $\le 4$
           &  {\ttbar}; QCD, $WW$, $WZ$            \\
    11     &  Min. missing transverse energy
           &  {\metc}, {\met}
           &  $> 15$~GeV
           &  QCD, $W$+jets                        \\
    12     &  Mismeasured {\met} (``triangle cuts'')
       & & &  QCD                                  \\
    13     &  Mismeasured tagging muon
           &  $p_T({\rm tag~} \mu)$
           &  $< 500$~GeV
           &  Mismeasured events
\vspace{0.04 in}                                   \\
\multicolumn{5}{l}{\underline{Muon Channel Only}}
\vspace{0.04 in}                                   \\
    14     &  Mismeasured isolated muon
           &  $p_T({\rm isol~} \mu)$, {\met}
           &  $< 250$~GeV
           &  Mismeasured events                   \\
    15     &  Mismeasured isolated muon
           &  $p_T({\rm isol~} \mu)$
           &  $< 500$~GeV
           &  Mismeasured events                   \\
    16     &  No back-to-back muons
           &  $\Delta\phi(\rm{isol~} \mu,\rm{tag~} \mu)$
           &  $< 2.4$~rad
           &  Cosmic rays
\vspace{0.04 in}                                   \\
\hline \hline
\end{tabular}
\caption[tab12]{The loose event selection variables and cutoffs in
the electron and muon channels.}
\label{table12}
\end{center}
\end{table}

\begin{table}[!h!tbp]
\begin{center}
\begin{tabular}{lcc}
\hline \hline
 & & \\ [-0.12 in]
\multicolumn{3}{c}{\large{Loose Selection Efficiencies}}
\vspace{0.03 in}                                 \\
\hline
Event Type  &  Electron Channel  &  Muon Channel \\
\hline
 & & \\ [-0.12 in]
Signals         &          &          \\
~~MC $tb$       &  $87\%$  &  $50\%$  \\
~~MC $tqb$      &  $83\%$  &  $53\%$
\vspace{0.02 in}                      \\
Backgrounds     &          &          \\
~~MC {\ttbar}   &  $46\%$  &  $28\%$  \\
~~MC {\Wbb}     &  $79\%$  &  $42\%$  \\
~~MC {\Wcc}     &  $79\%$  &  $41\%$  \\
~~MC $Wjj$      &  $72\%$  &  $32\%$  \\
~~MC $WW$       &  $69\%$  &  $44\%$  \\
~~MC $WZ$       &  $77\%$  &  $41\%$  \\
~~QCD data      &  $10\%$  &  $11\%$
\vspace{0.04 in}                      \\
Signal data     &  $18\%$  &   $7\%$
\vspace{0.04 in}                      \\
\hline \hline
\end{tabular}
\caption[tab13]{Percentages of the tagged baseline event samples which
remain after the loose selections.}
\label{table13}
\end{center}
\end{table}

\clearpage

\section{Tight Event Selections}

Table~\ref{table14} shows the tight event selection variables and
cutoffs. The variables have been chosen for highest sensitivity to
separate signals from backgrounds, and the cutoffs are optimized by
maximizing the significance of the signal significance. After the
tight selections, there remain 12 candidates in the electron channel
data, and 5 in the muon channel. The percentage of expected signal
increases to $3.8\%$ in the electron channel and $4.2\%$ in the muon
channel.

\begin{table}[!h!tbp]
\begin{center}
\begin{tabular}{ccrcc}
\hline \hline
 & & & & \\ [-0.12 in]
\multicolumn{5}{c} {\large{Tight Event Selections}}
\vspace{0.03 in}                                      \\
\hline
 Cut   &   &   &   &   Main Background                \\
  No.  & Variable Name &  \multicolumn{1}{c}{Variable Definition}
       &     Cutoff     &   Rejected               \\
\hline
 & & & & \\ [-0.12 in]
\multicolumn{5}{l}{\underline{Electron Channel}}
\vspace{0.04 in}                                               \\
     1     &  $H_T^{{\rm j}12e\nu}$
           &  $E_T({\rm jet1}) + E_T({\rm jet2}) + E_T(e) + {\met}$
           &  $> 125$~GeV
           &  $W$+jets                                         \\
     2     &  $H_T^{{\rm j}34^{\prime}}$
           &  $E_T({\rm jet3}) + 5 \times E_T({\rm jet4})$
           &  $< 47$~GeV
           &  {\ttbar}                                         \\
     3     &  $H_T^{{\rm j}1(4\nu)}$
           &  $E_T({\rm jet1}) + 4 \times {\met}$
           &  $> 155$~GeV
           &  QCD
\vspace{0.04 in}                                               \\
\multicolumn{5}{l}{\underline{Muon Channel}}
\vspace{0.04 in}                                               \\
     1     &  $H_T^{{\rm j}1234}$
           & $E_T({\rm j1})+E_T({\rm j2})+E_T({\rm j3})+E_T({\rm j4})$
           &  $> 70$~GeV
           &  $W$+jets                                         \\
     2     &  $H_T^{{\rm j}34^{\prime}}$
           &  $E_T({\rm jet3}) + 5 \times E_T({\rm jet4})$
           &  $< 47$~GeV
           &  {\ttbar}
\vspace{0.04 in}                      \\
\hline \hline
\end{tabular}
\caption[tab14]{The tight event selection variables and cutoffs in the
electron and muon channels.}
\label{table14}
\end{center}
\end{table}


\section{Properties of the Candidate Events}

Table~\ref{table15} shows averages of some properties of the electron
channel data and MC events after the loose event selections. Values
for the muon channel are similar. The averages have been calculated
for events after all correction factors have been applied.

\begin{table}[!h!tbp]
\centerline{
\begin{tabular}{lccccccccccc}
\hline \hline
 & & & & & & & & & & & \\ [-0.12 in]
\multicolumn{12}{c} {\large{Average Properties of the Candidate Events}}
\vspace{0.03 in}                                      \\
\hline
       &        &       & Electron &        & Jet1  & Jet2  & Jet3  & Tag$\mu$1
       & Tag$\mu$1 & $m_{\rm top}$ & $m_{\rm top}$ \\
 Event & No. of & No. of & $E_T$  & {\met} & $E_T$ & $E_T$ & $E_T$ & $p_T$
       & Jet-ID & ($e,\nu, j1$)   & ($e,\nu, j2$)  \\
 Type  & Jets   & $\mu$-Tags & [GeV] & [GeV] & [GeV] & [GeV] & [GeV] & [GeV]
       &        & [GeV] & [GeV] \\
\hline
 & & & & & & & & & & & \\ [-0.12 in]
Signals         &     &      &    &    &     &    &    &    &     &     &     \\
~~MC $tb$       & 2.1 & 1.04 & 47 & 54 &  85 & 47 & 16 & 17 & 1.4 & 206 & 154 \\
~~MC $tqb$      & 2.5 & 1.04 & 44 & 52 &  80 & 46 & 26 & 14 & 1.5 & 204 & 167
\vspace{0.02 in}                                                              \\
Backgrounds     &     &      &    &    &     &    &    &    &     &     &     \\
~~MC {\ttbar}   & 3.4 & 1.07 & 50 & 64 & 104 & 66 & 39 & 18 & 1.6 & 232 & 181 \\
~~MC {\Wbb}     & 2.0 & 1.04 & 47 & 47 &  64 & 33 &  9 & 13 & 1.3 & 186 & 140 \\
~~MC {\Wcc}     & 2.0 & 1.00 & 45 & 46 &  62 & 33 & 10 & 13 & 1.3 & 177 & 135 \\
~~MC $Wjj$      & 2.0 & 1.00 & 45 & 46 &  64 & 31 & ---&  6 & 1.2 & 170 & 157 \\
~~MC $WW$       & 2.5 & 1.00 & 41 & 53 &  72 & 37 & 21 & 10 & 1.4 & 195 & 140 \\
~~MC $WZ$       & 2.5 & 1.03 & 46 & 54 &  73 & 35 & 18 & 12 & 1.4 & 207 & 156 \\
~~$W$+jets data & 2.5 & ---  & 44 & 40 &  71 & 35 & 20 & ---& --- & 190 & 150 \\
~~QCD data      & 3.7 & 1.01 & 19 & 25 &  68 & 44 & 15 & 16 & 1.4 & 192 & 167
\vspace{0.04 in}                                                              \\
Signal data     & 2.9 & 1.00 & 49 & 43 &  78 & 37 & 26 & 12 & 1.3 & 195 & 162
\vspace{0.04 in}                                                              \\
\hline \hline
\end{tabular}}
\caption[tab15]{Average values of some variables of the tagged
electron channel events that pass the loose cuts.}
\label{table15}
\end{table}

\clearpage

The $s$-channel single top quark events have an average of 2.1 jets
reconstructed, whereas the $t$-channel events have 2.5 jets, showing
that the second $b$~jet is often not reconstructed as it has quite
low $p_T$ (naively, one might expect an average of $\sim$3.1
jets). Diboson backgrounds have similar jet multiplicity to
$t$-channel signals, and {\ttbar} and QCD events have significantly
more jets on average.

We are most likely to identify a second tagging muon in {\ttbar}
events ($\sim$7$\%$ of the tagged {\ttbar} events have two tagging
muons), since there are always two central, energetic $b$~jets.
However, the rate of double tags in single top events, at $\sim$4$\%$
of the single tagged events, is still usefully higher than for other
backgrounds.

The kinematic properties of single top quark events lie between those
of the energetic {\ttbar} background events, and the more numerous
$W$+jets events. Thus, it is more difficult to separate single top
signals from background than to identify {\ttbar} events.

\begin{table}[!h!tbp]
\centerline{
\begin{tabular}{cccccccccccc}
\hline \hline
 & & & & & & & & & & & \\ [-0.12 in]
\multicolumn{12}{c} {\large{The Tagged Candidate Events -- Part 1 -- Main Properties}}
\vspace{0.03 in}                                                                                                          \\
\hline
         &      &      &        &         &         &        & Tight  & Tight  & Tight  & $m_{\rm top}$  & $m_{\rm top}$  \\
  Event  & Cut  & Run  & Event  & No. of  & Tagged  & Event  & Var.1  & Var.2  & Var.3  & ($l,\nu, j1$)  & ($l,\nu, j2$)  \\
  No.    & Set  & No.  & No.    & Jets    & Jet No. & Zone   & [GeV]  & [GeV]  & [GeV]  & [GeV]          & [GeV]          \\
\hline
 & & & & & & & & & & & \\ [-0.12 in]
\multicolumn{12}{c}{\underline{Electron Channel}}
\vspace{0.04 in}                                                                                                                    \\
{\bf  1} &{\bf T} &{\bf 76339} &{\bf 11474} &{\bf  2} &{\bf 1} &{\bf EC/CF1} &{\bf 199} &{\bf   0} &{\bf 254} &{\bf 209} &{\bf 156} \\
{\bf  2} &{\bf T} &{\bf 76579} &{\bf 30349} &{\bf  2} &{\bf 1} &{\bf CC/CF1} &{\bf 144} &{\bf   0} &{\bf 176} &{\bf 138} &{\bf 358} \\
{\bf  3} &{\bf T} &{\bf 81336} &{\bf  3313} &{\bf  3} &{\bf 1} &{\bf CC/CF2} &{\bf 223} &{\bf   9} &{\bf 407} &{\bf 228} &{\bf 161} \\
{\bf  4} &{\bf T} &{\bf 83044} &{\bf  8929} &{\bf  2} &{\bf 1} &{\bf CC/CF1} &{\bf 155} &{\bf   0} &{\bf 232} &{\bf 155} &{\bf 155} \\
{\bf  5} &{\bf T} &{\bf 84225} &{\bf 12800} &{\bf  2} &{\bf 1} &{\bf CC/CF1} &{\bf 149} &{\bf   0} &{\bf 168} &{\bf 139} &{\bf 138} \\
{\bf  6} &{\bf T} &{\bf 84681} &{\bf 13015} &{\bf  3} &{\bf 1} &{\bf CC/CF2} &{\bf 255} &{\bf  22} &{\bf 232} &{\bf 178} &{\bf 167} \\
{\bf  7} &{\bf T} &{\bf 84998} &{\bf 15724} &{\bf  2} &{\bf 2} &{\bf CC/CF1} &{\bf 226} &{\bf   0} &{\bf 398} &{\bf 210} &{\bf 119} \\
{\bf  8} &{\bf T} &{\bf 85780} &{\bf 17023} &{\bf  3} &{\bf 1} &{\bf CC/CF1} &{\bf 199} &{\bf  21} &{\bf 295} &{\bf 185} &{\bf 113} \\
{\bf  9} &{\bf T} &{\bf 85781} &{\bf 10705} &{\bf  3} &{\bf 2} &{\bf CC/CF1} &{\bf 194} &{\bf  31} &{\bf 295} &{\bf 196} &{\bf 127} \\
{\bf 10} &{\bf T} &{\bf 87987} &{\bf  1228} &{\bf  3} &{\bf 1} &{\bf CC/CF1} &{\bf 225} &{\bf  22} &{\bf 209} &{\bf 161} &{\bf 213} \\
{\bf 11} &{\bf T} &{\bf 91981} &{\bf 29111} &{\bf  2} &{\bf 1} &{\bf CC/CF1} &{\bf 282} &{\bf   0} &{\bf 243} &{\bf 248} &{\bf 171} \\
{\bf 12} &{\bf T} &{\bf 93039} &{\bf 29631} &{\bf  3} &{\bf 1} &{\bf EC/CF1} &{\bf 181} &{\bf  21} &{\bf 366} &{\bf 211} &{\bf 157} \\
     13  &     L  &     63799  &     13414  &      3  &     2  &     CC/CF1  &{\it 111} &      20  &{\it 145} &     133  &     123  \\
     14  &     L  &     85437  &     31896  &      3  &     1  &     CC/CF1  &{\it 124} &       6  &{\it 139} &     120  &     111  \\
     15  &     L  &     87449  &      1860  &      4  &     1  &     EC/CF1  &     130  &{\it  50} &{\it 139} &     165  &     140  \\
     16  &     L  &     88610  &      9826  &      4  &     3  &     CC/CF1  &     191  &{\it  48} &     322  &     205  &     169  \\
     17  &     L  &     89372  &     12467  &      4  &     3  &     CC/CF1  &     428  &{\it 187} &     352  &     382  &     271  \\
     18  &     L  &     89546  &      9435  &      2  &     1  &     CC/CF1  &{\it 104} &       0  &{\it 110} &     122  &     127  \\
     19  &     L  &     89708  &     34735  &      3  &     1  &     CC/CF2  &     130  &      13  &{\it 128} &     168  &      96  \\
     20  &     L  &     91206  &     13727  &      3  &     1  &     CC/CF2  &     449  &{\it  51} &     458  &     363  &     133  \\
     21  &     L  &     92225  &     17428  &      4  &     1  &     CC/CF1  &     231  &{\it 241} &{\it 146} &     172  &     205
\vspace{0.04 in}                                                                                                                    \\
\hline
 & & & & & & & & & & & \\ [-0.12 in]
\multicolumn{12}{c}{\underline{Muon Channel}}
\vspace{0.04 in}                                                                                                                    \\
{\bf  1} &{\bf T} &{\bf 81693} &{\bf 11454} &{\bf  3} &{\bf 2} &{\bf CF1/CF2}&{\bf 134} & {\bf 16} &          &{\bf 214} &{\bf  97} \\
{\bf  2} &{\bf T} &{\bf 83077} &{\bf  9934} &{\bf  3}&{\bf 2/2}&{\bf CF1/CF1}&{\bf 243} & {\bf 39} &          &{\bf 177} &{\bf 164} \\
{\bf  3} &{\bf T} &{\bf 83078} &{\bf 15303} &{\bf  3} &{\bf 2} &{\bf CF1/CF1}&{\bf  86} & {\bf 16} &          &{\bf 121} &{\bf 112} \\
{\bf  4} &{\bf T} &{\bf 84695} &{\bf 29699} &{\bf  3} &{\bf 2} &{\bf CF1/CF1}&{\bf 241} & {\bf 36} &          &{\bf 185} &{\bf 174} \\
{\bf  5} &{\bf T} &{\bf 90572} &{\bf 46085} &{\bf  2} &{\bf 2} &{\bf EF/CF1} &{\bf 207} & {\bf  0} &          &{\bf 304} &{\bf 253} \\
      6  &     L  &     85735  &     22588  &      2  &     2  &     CF1/CF1 &{\it  58} &       0  &          &     135  &     118  \\
      7  &     L  &     87882  &     17098  &      4  &     3  &     CF1/CF1 &     165  & {\it 103}&          &     154  &     175  \\
      8  &     L  &     92238  &        29  &      2  &     1  &     EF/CF2  &{\it  69} &       0  &          &     167  &     121   
\vspace{0.04 in}                                                                                                                    \\
\hline \hline
\end{tabular}}
\caption[tab16]{Main properties of the tagged candidate events after
the loose event selection cuts. Values in bold type are for events
which also pass the tight cuts. Values in italics show where the loose
cut-set events fail one or more of the tight cuts.}
\label{table16}
\end{table}

\clearpage

Tables~\ref{table16} and \ref{table17} show the properties of the
individual data candidates that pass the loose or tight event
selections. The event zone abbreviations in Table~\ref{table16} are
defined as follows: CC = electron in central calorimeter (${\mdeta} <
1.1$); EC = electron in an end calorimeter ($1.5 < {\mdeta} < 2.5$);
CF1 = muon in central spectrometer (${\mdeta} \leq 0.6$); CF2 = muon
in central spectrometer (${\mdeta} > 0.6$); EF = muon in end
spectrometer (${\mdeta} < 1.7$). The junction between the central and
end regions of the muon spectrometer occurs between $0.8 < {\mdeta} <
1.2$, depending on $\phi$.

\begin{table}[!h!tbp]
\begin{center}
\begin{tabular}{ccccccccccc}
\hline \hline
 & & & & & & & & & & \\ [-0.12 in]
\multicolumn{11}{c} {\large{The Tagged Candidate Events -- Part 2 -- Kinematics}}
\vspace{0.03 in}                                                                                                \\
\hline
 Event  & Lepton  & Lepton   &         & Jet 1  & Jet 1    & Jet 2  & Jet 2    & Jet 3  & Jet 3    & Tag $\mu$  \\
 No.    & $E_T$   & {\deta}  & {\met}  & $E_T$  & {\deta}  & $E_T$  & {\deta}  & $E_T$  & {\deta}  & $p_T$      \\
        & [GeV]   &          & [GeV]   & [GeV]  &          & [GeV]  &          & [GeV]  &          & [GeV]      \\
\hline
 & & & & & & & & & & \\ [-0.12 in]
\multicolumn{11}{c}{\underline{Electron Channel}}
\vspace{0.04 in}                                                                                                             \\
{\bf  1}  &{\bf  43} &{\bf$-1.7$} &{\bf 41} &{\bf  88} &{\bf$-0.0$} &{\bf 26} &{\bf$ 1.0$} &{\bf ---} &{\bf  --- } &{\bf  6} \\
{\bf  2}  &{\bf  21} &{\bf$-0.3$} &{\bf 31} &{\bf  52} &{\bf$ 0.2$} &{\bf 40} &{\bf$-2.6$} &{\bf ---} &{\bf  --- } &{\bf 26} \\
{\bf  3}  &{\bf  24} &{\bf$ 1.0$} &{\bf 78} &{\bf  97} &{\bf$ 0.5$} &{\bf 25} &{\bf$ 2.0$} &{\bf   9} &{\bf$-0.2$} &{\bf  5} \\
{\bf  4}  &{\bf  33} &{\bf$-0.3$} &{\bf 43} &{\bf  60} &{\bf$ 0.5$} &{\bf 19} &{\bf$-1.7$} &{\bf ---} &{\bf  --- } &{\bf  6} \\
{\bf  5}  &{\bf  63} &{\bf$ 0.1$} &{\bf 33} &{\bf  37} &{\bf$-0.6$} &{\bf 17} &{\bf$-1.8$} &{\bf ---} &{\bf  --- } &{\bf  8} \\
{\bf  6}  &{\bf  61} &{\bf$-0.9$} &{\bf 35} &{\bf  94} &{\bf$-0.6$} &{\bf 65} &{\bf$ 0.3$} &{\bf  22} &{\bf$ 1.7$} &{\bf 21} \\
{\bf  7}  &{\bf  26} &{\bf$-0.7$} &{\bf 77} &{\bf  88} &{\bf$-0.9$} &{\bf 35} &{\bf$-0.3$} &{\bf ---} &{\bf  --- } &{\bf  7} \\
{\bf  8}  &{\bf  38} &{\bf$-0.5$} &{\bf 52} &{\bf  87} &{\bf$ 0.0$} &{\bf 22} &{\bf$-0.6$} &{\bf  21} &{\bf$-1.5$} &{\bf  9} \\
{\bf  9}  &{\bf  22} &{\bf$-0.0$} &{\bf 52} &{\bf  86} &{\bf$ 0.7$} &{\bf 34} &{\bf$ 0.2$} &{\bf  31} &{\bf$-1.8$} &{\bf  6} \\
{\bf 10}  &{\bf  66} &{\bf$ 0.7$} &{\bf 35} &{\bf  68} &{\bf$ 0.3$} &{\bf 55} &{\bf$-1.3$} &{\bf  22} &{\bf$ 0.4$} &{\bf 15} \\
{\bf 11}  &{\bf  80} &{\bf$-0.6$} &{\bf 34} &{\bf 107} &{\bf$ 0.3$} &{\bf 61} &{\bf$-0.4$} &{\bf ---} &{\bf  --- } &{\bf 25} \\
{\bf 12}  &{\bf  28} &{\bf$-2.2$} &{\bf 79} &{\bf  49} &{\bf$ 0.1$} &{\bf 24} &{\bf$ 0.3$} &{\bf  21} &{\bf$-0.9$} &{\bf  5} \\
     13   &      20  &    $-0.8$  &     27  &      37  &    $ 1.2$  &     27  &    $-0.5$  &      20  &    $ 0.1$  &     12  \\
     14   &      41  &    $ 0.8$  &     26  &      35  &    $-0.1$  &     22  &    $ 0.6$  &       6  &    $ 0.1$  &      5  \\
     15   &      26  &    $ 2.0$  &     19  &      65  &    $-0.2$  &     21  &    $-1.3$  &      16  &    $ 2.6$  &     14  \\
     16   &      65  &    $-0.9$  &     72  &      33  &    $ 1.0$  &     21  &    $-0.4$  &      20  &    $ 0.2$  &      6  \\
     17   &     124  &    $ 0.9$  &     36  &     209  &    $ 0.6$  &     59  &    $-1.1$  &      52  &    $ 0.4$  &     17  \\
     18   &      35  &    $-0.7$  &     20  &      28  &    $ 0.3$  &     21  &    $ 1.4$  &     ---  &      ---   &      7  \\
     19   &      35  &    $ 1.0$  &     17  &      60  &    $-0.9$  &     18  &    $-0.3$  &      13  &    $-3.2$  &     14  \\
     20   &      91  &    $-0.9$  &     65  &     196  &    $-0.8$  &     97  &    $-1.1$  &      51  &    $-0.1$  &     35  \\
     21   &      83  &    $ 0.6$  &     20  &      67  &    $-0.2$  &     61  &    $-0.9$  &      57  &    $-0.2$  &      9
\vspace{0.04 in}                                                                                                             \\
\hline                                                                                                                      
 & & & & & & & & & & \\ [-0.12 in]
\multicolumn{11}{c}{\underline{Muon Channel}}
\vspace{0.04 in}                                                                                                             \\
{\bf  1}  &{\bf  38} &{\bf$ 0.1$} &{\bf 52} &{\bf  87} &{\bf$ 1.5$} &{\bf 31} &{\bf$ 0.7$} &{\bf  16} &{\bf$ 0.6$} &{\bf  7} \\
{\bf  2}  &{\bf  36} &{\bf$ 0.7$} &{\bf 19} &{\bf 114} &{\bf$-0.9$} &{\bf 90} &{\bf$ 0.7$} &{\bf  39} &{\bf$ 1.7$} &{\bf  8} \\
{\bf  3}  &{\bf  27} &{\bf$ 0.1$} &{\bf 20} &{\bf  43} &{\bf$-0.6$} &{\bf 26} &{\bf$-0.1$} &{\bf  16} &{\bf$ 0.7$} &{\bf  5} \\
{\bf  4}  &{\bf  47} &{\bf$-0.1$} &{\bf 38} &{\bf 115} &{\bf$-0.8$} &{\bf 89} &{\bf$-0.5$} &{\bf  36} &{\bf$ 1.0$} &{\bf 33} \\
{\bf  5}  &{\bf  70} &{\bf$-1.3$} &{\bf 83} &{\bf 132} &{\bf$-0.6$} &{\bf 75} &{\bf$ 0.0$} &{\bf ---} &{\bf  --- } &{\bf 18} \\
      6   &      24  &    $ 0.3$  &     69  &      33  &    $ 0.7$  &     25  &    $-0.5$  &     ---  &      ---   &      5  \\
      7   &      22  &    $-0.1$  &     16  &      73  &    $ 1.1$  &     49  &    $-0.6$  &      28  &    $-0.1$  &      6  \\
      8   &      21  &    $ 1.3$  &     51  &      51  &    $-1.0$  &     18  &    $ 2.1$  &     ---  &      ---   &    110 
\vspace{0.04 in}                                                                                                             \\
\hline \hline
\end{tabular}
\caption[tab17]{Kinematic properties of the tagged candidate events
after the loose event selections. Values in bold type are for events
which also pass the tight cuts. Not shown are electron event~15's
jet~4, which has $E_T$ = 7~GeV and {\deta} = 0.2, electron event~16's
jet~4, which has $E_T$ = 6~GeV and {\deta} = $-3.3$, electron
event~17's jet~4, which has $E_T$ = 27~GeV and {\deta} = 1.1, and
electron event~21's jet~4, which has $E_T$ = 37~GeV and {\deta} =
$-0.8$. Also not shown are muon event~2's second tagging muon (jet~2
is double-tagged), which has $p_T$ = 4~GeV, and muon event~7's jet~4,
which has $E_T$ = 15~GeV, and {\deta} = $-2.9$.}
\label{table17}
\end{center}
\end{table}

\clearpage

\section{Summary}

This paper has presented details of the baseline and loose event
selections used in {\dzero}'s search for single top quark production
at the Tevatron collider. The baseline criteria are chosen to be
ultra-loose, to keep maximal signal acceptance. The loose criteria are
chosen to remove mismeasured events from the samples and to reject
events which are obviously not signals. Application of these
selections reduces the data sample from approximately one million
events in each of the electron and muon channels to 21 {\ejm}
candidates and 8 {\mjm} candidates. The combined $s$-channel and
$t$-channel signal acceptance is $2.6\%$ in the electron channel
before requiring a tagging muon to identify a $b$~jet, and $0.22\%$
after this requirement. In the muon channel, the combined acceptance
is $1.8\%$ before tagging and $0.11\%$ after. These acceptances are
percentages of the total single top quark cross section with no
branching fractions included. Most of the acceptance is lost by the
demand for a fiducial isolated lepton that passes strict particle
identification criteria, part of the baseline selections. The jet
$E_T$ thresholds, {\met} threshold, and mismeasured {\met} cuts make
up the rest of the inefficiency in the electron channel. In the muon
channel, an additional significant loss occurs from the cut applied to
reject cosmic ray contamination.

We have presented averages of various properties of the signal and
background samples after the loose selections. This information can
help to determine how best to separate the samples. We have also shown
the detailed properties of the candidate events remaining in the data
after the loose selections. There is one double-tagged candidate, a
muon channel event with two tagging muons in the second-highest-$E_T$
jet. A double-tagged jet is highly likely to be a $b$~jet, thus making
this event particularly interesting. The reconstructed invariant mass
from the isolated muon, missing transverse energy (interpreted as a
neutrino from a $W$~boson decay) and either of the first two jets
($m_{l\nu j} = 177$~GeV with jet 1 and 164~GeV with jet 2) is quite
close to the average Tevatron value of the top quark mass,
174.3~GeV. In Run~2, we hope to have a much larger sample of data,
with far more efficient $b$~jet tagging using a silicon vertex
detector as well as lepton tagging, and thereby to observe many more
such candidates.



\begin{thebibliography}{99}

\bibitem{sintop-prd}
B.~Abbott {\it et al.}, {\dzero} Collaboration, {\it ``Search for
Electroweak Production of Single Top Quarks in {\ppbar} Collisions''},
Phys. Rev. D {\bf 63}, 031101(R) (2001).

\bibitem{cdfpaper}
P.~Savard, {``\it Search for Single Top Production with CDF''}, to
appear in the proceedings of the Meeting of the Division of Particles
and Fields of the American Physical Society, The Ohio State
University, Columbus, OH, August 2000.

\bibitem{willenbrock-s}
M.C.~Smith and S.~Willenbrock,{``\it QCD and Yukawa Corrections to
Single Top Quark Production via ${\qqbar}{\rar} t{\bbar}$''},
Phys. Rev. D {\bf 54}, 6696 (1996).

This calculation has been recently updated by Brian Harris to use
CTEQ5M1 as the parton distribution function set (PDF) instead of
CTEQ3M, for a top quark mass $m_t=174.3$~GeV, the Tevatron average
value. The PDF change increases the result by $2.5\%$. The quoted
error includes contributions from the choice of scale and PDF, and
from the 5.1~GeV error on $m_t$, which dominates.

\bibitem{willenbrock-t}
T.~Stelzer, Z.~Sullivan, and S.~Willenbrock,{``\it Single Top Quark
Production via $W$-Gluon Fusion at Next-to-Leading Order''},
Phys. Rev. D {\bf 56}, 5919 (1997).

This calculation has been recently updated by Zack Sullivan to use
CTEQ5M1 instead of CTEQ3M, for a top quark mass of 174.3~GeV. The new
CTEQ PDF has a major bug fixed in the evolution equations that
generate the $b$~quark distributions (present in all previous CTEQ and
MRS PDFs). The bug fix lowers the $t$-channel single top quark cross
section by $15\%$. The quoted error includes contributions from the
choice of scale and PDF, and from the error on $m_t$.

\bibitem{heinson-dpf}
A.P.~Heinson, {``\it Search for Electroweak Production of Single Top
Quarks at {\dzero}''}, to appear in the proceedings of the Meeting of
the Division of Particles and Fields of the American Physical Society,
The Ohio State University, Columbus, OH, August 2000.

\end{thebibliography}
\end{document}